\documentclass[12pt]{iopart}
\usepackage{natbib}
\usepackage{graphicx} 
\usepackage{iopams}  
\begin{document}

\title[Cosmogenic Neutrinos: parameter space and detectabilty from PeV to ZeV]{Cosmogenic Neutrinos: \\parameter space and detectabilty from PeV to ZeV}

\author{K. Kotera$^1$, D. Allard$^2$ and A. V. Olinto$^1$}

\address{$^1$Department of Astronomy \& Astrophysics, Enrico Fermi Institute,
and Kavli Institute for Cosmological Physics, The University of Chicago, Chicago, Illinois 60637, USA.\\
$^2$Laboratoire Astroparticules et Cosmologie  (APC), Universit\'e Paris 7/CNRS, 10 rue A. Domon et L. Duquet, 75205 Paris Cedex 13, France.}
\ead{kotera@uchicago.edu}
\begin{abstract}
While propagating from their source to the observer, ultrahigh energy cosmic rays interact with cosmological photon backgrounds and generate to the so-called ``cosmogenic neutrinos". Here we study the parameter space of the cosmogenic neutrino flux given recent cosmic ray data and updates on plausible source evolution models. The shape and normalization of the cosmogenic neutrino flux are very sensitive to some of the current unknowns of ultrahigh energy cosmic ray sources and composition. We investigate various chemical compositions and maximum proton acceleration energies $E_{p, \rm max}$ which are allowed by current observations. We consider different models of source evolution in redshift and three possible scenarios for the Galactic to extragalactic transition. 

We summarize the parameter space for cosmogenic neutrinos into three regions: an optimistic scenario that is currently being constrained by observations, a plausible range of models in which we base many of our rate estimates, and a pessimistic scenario that will postpone detection for decades to come. We present the implications of these three scenarios for the detection of cosmogenic neutrinos from PeV to ZeV ($10^{14-21}$~eV) with the existing and upcoming instruments. In the plausible range of parameters, the narrow flux variability in the EeV energy region assures low but detectable rates for IceCube   ($0.06-0.2$ neutrino per year) and the Pierre Auger Observatory ($0.03-0.06$ neutrino per year), and detection should happen in the next decade. If EeV neutrinos are detected, PeV information can help select between competing models of cosmic ray composition at the highest energy and the Galactic to extragalactic transition at ankle energies.  With improved sensitivity, ZeV neutrino observatories, such as ANITA and JEM-EUSO could explore and place limits on the maximum acceleration energy. 
\end{abstract}

\maketitle

\section{Introduction}

The idea that a ``guaranteed'' flux of detectable extragalactic neutrinos should be produced by the propagation of ultrahigh energy cosmic rays as they interact with the ambient photon backgrounds
 \citep{BZ69, Stecker79} 
 has encouraged efforts to detect them for decades (see, e.g.,  \citealp{AM09}). One important assumption, that cosmic rays are extragalactic  at the highest energy, has been verified by the detection of the Greisen-Zatsepin-Kuzmin (GZK) cutoff feature \citep{G66,ZK66} in the cosmic ray spectrum \citep{Abbasi09,Auger08} 
and by the indication of anisotropies in the cosmic ray sky distribution at the highest energies \citep{Auger1,Auger2}. 
These findings herald the possibility of detecting ultrahigh energy neutrinos in the near future and  a possible resolution to the mystery behind the origin of ultrahigh energy cosmic rays (UHECRs). 

At present the sources of ultrahigh energy cosmic rays, their location, cosmological evolution, and maximum energy, as well as the injected composition, remain unknown. A multi-messenger approach with the detection of secondary neutrinos and photons can lead to a resolution of this quest for sources. At the highest energies neutrinos are particularly useful because, unlike cosmic-rays and photons, they are not absorbed by the cosmic backgrounds while propagating through the Universe. These so-called ``cosmogenic neutrinos" bear the information of ultrahigh energy cosmic ray sources up to high redshifts, and can help constrain their nature, injection spectrum, distribution, and evolution.

A number of authors have estimated the cosmogenic flux with varying assumptions (e.g., \citealp{ESS01,Ave05,Seckel05,HTS05,Berezinsky06,Stanev06,Allard06,Takami09}, and see \citealp{AM09} for a recent review): in particular, \cite{ESS01} and \cite{Ave05} calculated analytically and semi-analytically the expected neutrino flux around PeV energies, considering cosmic ray interactions with CMB photons and a homogeneous source distribution. They find that IceCube could detect $\sim 0.2$ events above $E_\nu\gtrsim 1~$PeV. \cite{Allard06} made a detailed numerical study, including the influence of infrared (IR) and ultraviolet (UV) photon backgrounds and exploring various cosmic ray chemical compositions and source evolution models. \cite{Takami09} explored the influence of different source evolution and transition models in the case cosmic rays are protons. They highlighted the robustness of the EeV neutrino flux and the considerable variation of the PeV flux according to transition models. 

In this work, we explore the entire parameter space of cosmogenic neutrinos allowed by ultrahigh energy cosmic ray observations. We review and update the parameters used in previous work and estimate detection rates in the light of the recent experimental progress. 

The existing and upcoming high energy neutrino detectors roughly cover three energy ranges: PeV (= $10^{15}$ eV), EeV (= $10^{18}$ eV), and ZeV (= $10^{21}$ eV).  ANTARES, IceCube, and the future KM3Net are large water or ice cubic detectors that aim at observing events around PeV energies. Once completed, IceCube will also have a very good sensitivity at higher energies, and will ultimately be able to cover a wide energy range from about 1~PeV to $\sim10$~EeV.  Experiments primarily dedicated to the detection of cosmic rays like the Pierre Auger Observatory and the Telescope Array have their best neutrino sensitivities in the EeV energy range.  The radio telescope ANITA and the fluorescence telescope JEM-EUSO are most effective at the highest energy neutrinos around 0.1 ZeV. 

We demonstrate in this paper that the detection of cosmogenic neutrinos is challenging for current detectors, unless cosmic ray sources fall into the optimistic category. If neutrinos are observed, each energy range would enable one to explore the influence of a specific set of cosmic ray source parameters. 
In section~\ref{section:fluxes}, we examine the effects of the major unknown that concern ultrahigh energy cosmic ray sources: the injected chemical composition, the maximum proton acceleration energy, source evolution models, and outstanding scenarios for the Galactic to extragalactic transition. Section~\ref{section:detection} discusses the implication of our calculations in terms of detection by current and upcoming neutrino experiments. We focus alternately on the three main energy ranges covered by these instruments and conclude in section~\ref{section:conclusion}.

\section{Neutrino fluxes}\label{section:fluxes}

We calculate the fluxes of neutrinos generated by the propagation of ultrahigh energy cosmic rays over cosmological distances. For this purpose we use a complete numerical Monte Carlo method that takes into account the interactions of nuclei with cosmic background radiations (see \citealp{Allard05} and \citealp{KAM09} for more details). Our modeling of the background radiations includes redshift evolutions of the cosmic microwave background (CMB) and of the diffuse infrared, optical, and ultraviolet  backgrounds modeled according to \cite{SMS06}, hereafter referred to as the IR/UV background.  The IR/UV background and its cosmological evolution are not as well known as the CMB, though recent measurements (see discussion in \citealp{SMS06} and \citealp{Allard06}) have lead to better constraints. One can notice that \cite{KBMH04} find higher photon densities in the far infrared bump at large redshift compared to \cite{SMS06}, and lower densities in the optical and UV range. The latter difference implies neutrino fluxes a factor of two higher in our study compared to \cite{Takami09} around $3\times10^{15}$ eV, and the agreement becomes very good above $10^{16}$~eV. The dispersion of neutrino fluxes due to the astrophysical hypotheses we consider in the following is  far larger than the one introduced by different modelings of the IR/UV backgrounds. The latter do not have a large impact on the discussion of the detectability of cosmogenic neutrinos in the PeV region.

Particles are injected with energy between $E_{\rm min}=10^{16}$~eV and $Z\times E_{p,\rm max}$, following a power-law spectrum: ${\rm d}N/{\rm d}E\propto E^{-\alpha}$, with an exponential cut-off above $Z\times E_{p,\rm max}$. Apart from section~\ref{subsection:epmax},  we use a fiducial maximum proton acceleration energy of $E_{p,\rm max}=10^{20.5}$~eV. Cosmic rays are followed down to energy $E_{\rm cr}=10^{15}~$eV where the interaction probability becomes negligible. The redshifts of their sources are distributed between $z=10^{-3}$  and $z=8$ and are weighted according to source evolution models.

In the following, we explore different source evolution and transition models, chemical compositions and maximum acceleration energies. For each scenario, the spectral index and the overall normalization of the cosmic ray flux is chosen to best fit the Auger data. These combined parameters lead to a particular shape and normalization of the produced cosmogenic neutrino fluxes. Table~\ref{table:indices} presents the spectral indices corresponding to each model.

\Table{\label{table:indices}Spectral indices corresponding to models presented in section~\ref{section:fluxes}. Abreviations for source evolution models are defined in section~\ref{subsection:source_evol}. Galactic to extragalactic transition models are described in section~\ref{subsection:transition} and chemical compositions in section~\ref{subsection:composition}. }
\br
  composition &  transition &source evolution & $\alpha$\\
\mr
pure protons 		& dip 		& uniform		& 2.6 \\
"			& "	 		& SFR1		& 2.5 \\
"			& "			& SFR2		& 2.5 \\
"			& "			& GRB1		& 2.4 \\
"			& "			& GRB2		& 2.4 \\
"			& "			& FRII		& 2.3 \\
"			& WW		& SFR1		& 2.1 \\
Galactic mix		& Galactic mix		& SFR1		& 2.1 \\
pure iron			& Galactic mix		& SFR1		& 2.0 \\
iron rich, low $E_{\rm p, max}$& Galactic mix & SFR1	& 1.2\\
\br
\endTable

\begin{figure}[htb]
\begin{center}
\includegraphics[width=0.6\textwidth]{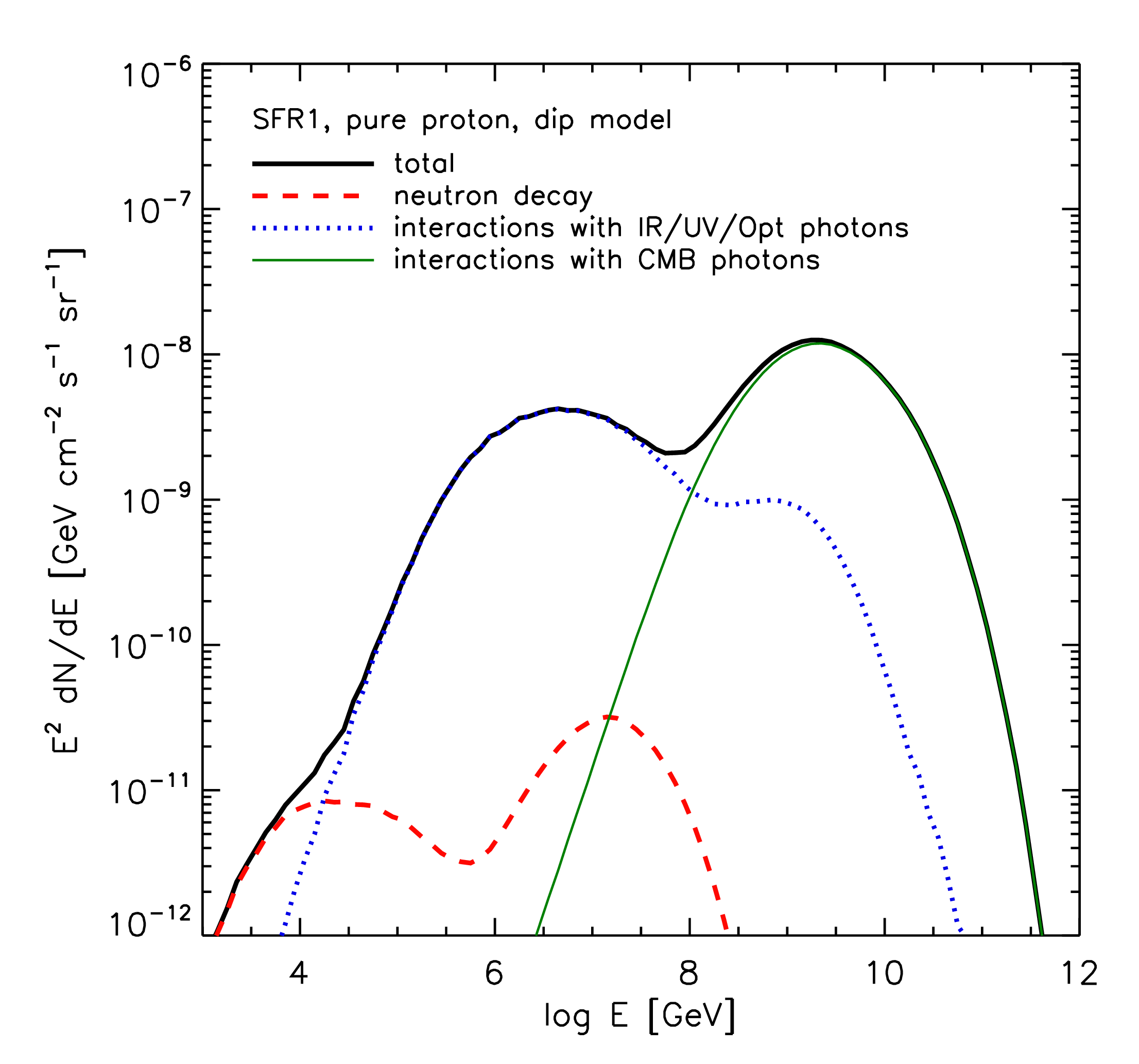} 
\caption{Contribution of the different processes to the neutrino flux, considering all flavors. The case of a pure proton composition, assuming a star formation rate type evolution for the source emissivity \citep{HB06} and a dip transition model \citep{BGG06} is presented. The black solid line indicates the total flux. The green solid line represents the neutrino emission due to the interaction of cosmic rays with CMB photons and the blue dotted line with UV, optical, and IR  photons. The red dashed line is the contribution of the neutron decay (neutrons are produced through photo-hadronic interactions).}  
\label{fig:basic}
\end{center}
\end{figure}

Let us note that our cosmogenic neutrino fluxes all present the same basic shape that can be understood from Fig.~\ref{fig:basic}. Neutrinos are produced via two principal channels: by pion decay or by neutron decay. The contribution of the latter channel is plotted in red dashed line in Fig.~\ref{fig:basic}. Pions are produced through the interaction of neutrons, protons, and heavier nuclei with the CMB and the IR/UV radiation. Those two backgrounds are responsible for the presence of the two bumps around a few PeV and a few EeV in the neutrino flux. We will see in the following that the height of the low energy bump mainly depends on the value of the injection spectral index, $\alpha$, corresponding to the various models. The high energy bump remains fairly independent of $\alpha$ and of transition models, depending mostly on the composition and on the maximum acceleration energy.

\subsection{Effets of different source evolutions}\label{subsection:source_evol}

\begin{figure*}[htb]
\begin{center}
\includegraphics[width=0.6\textwidth]{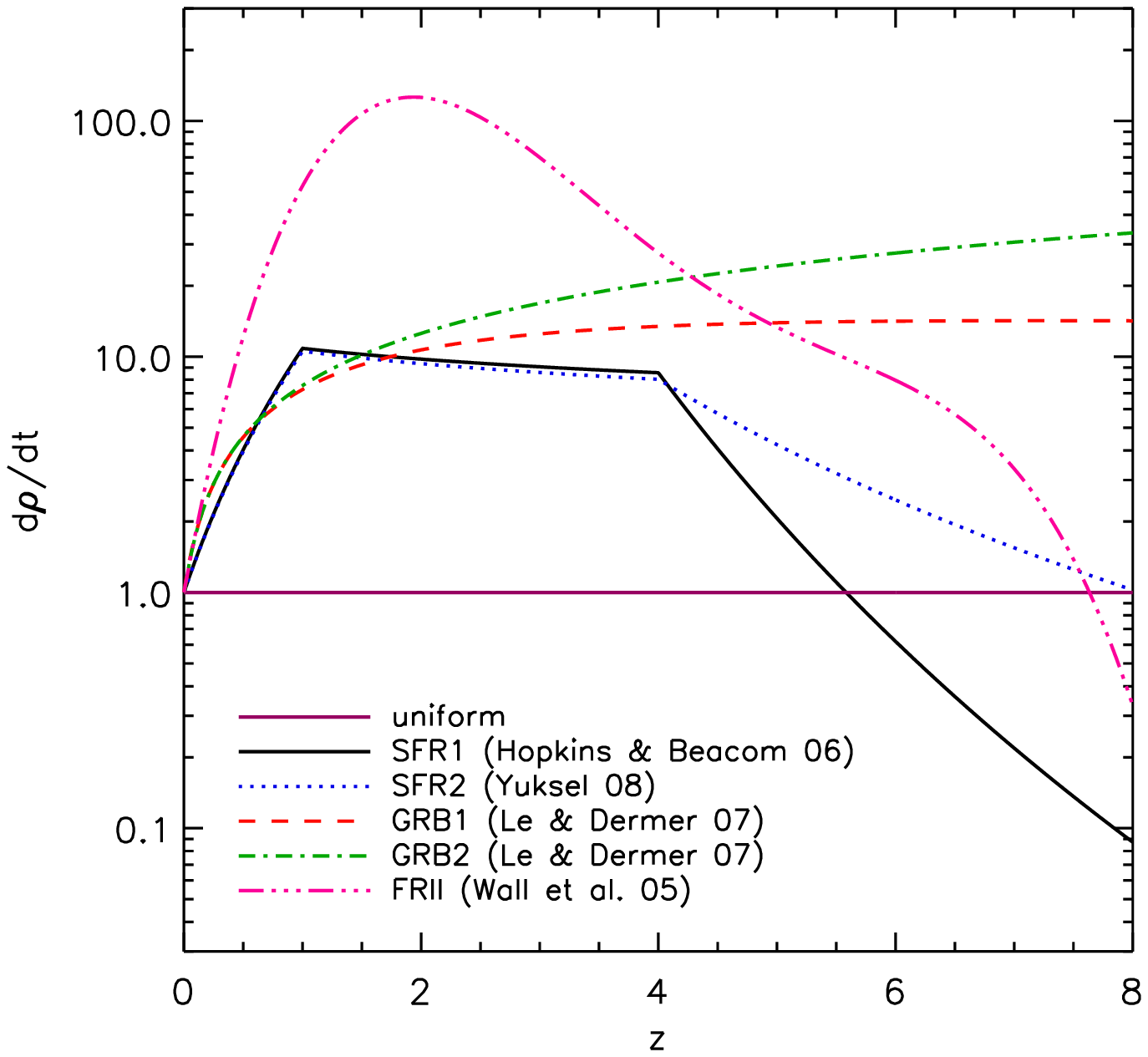} 
\includegraphics[width=0.6\textwidth]{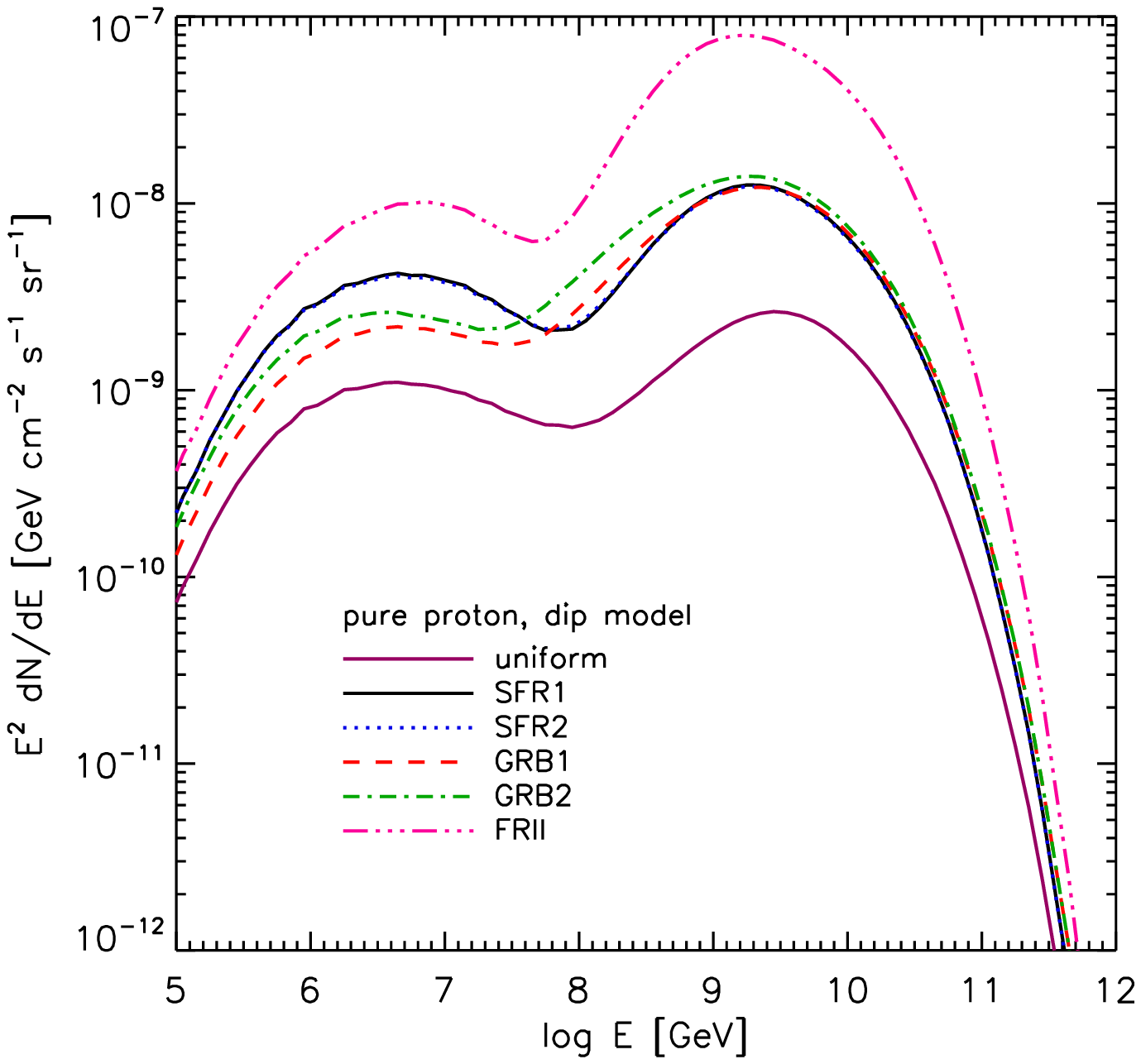} 
\caption{Top: source emissivity evolution with redshift, normalized to unity at $z=0$, for our six models described in the text. Bottom: effects of source evolution on neutrino fluxes for all flavors. We assume here a pure proton composition and a dip transition model. }  \label{fig:source_evol}
\end{center}
\end{figure*}

\begin{figure}[htb]
\begin{center}
\includegraphics[width=0.6\textwidth]{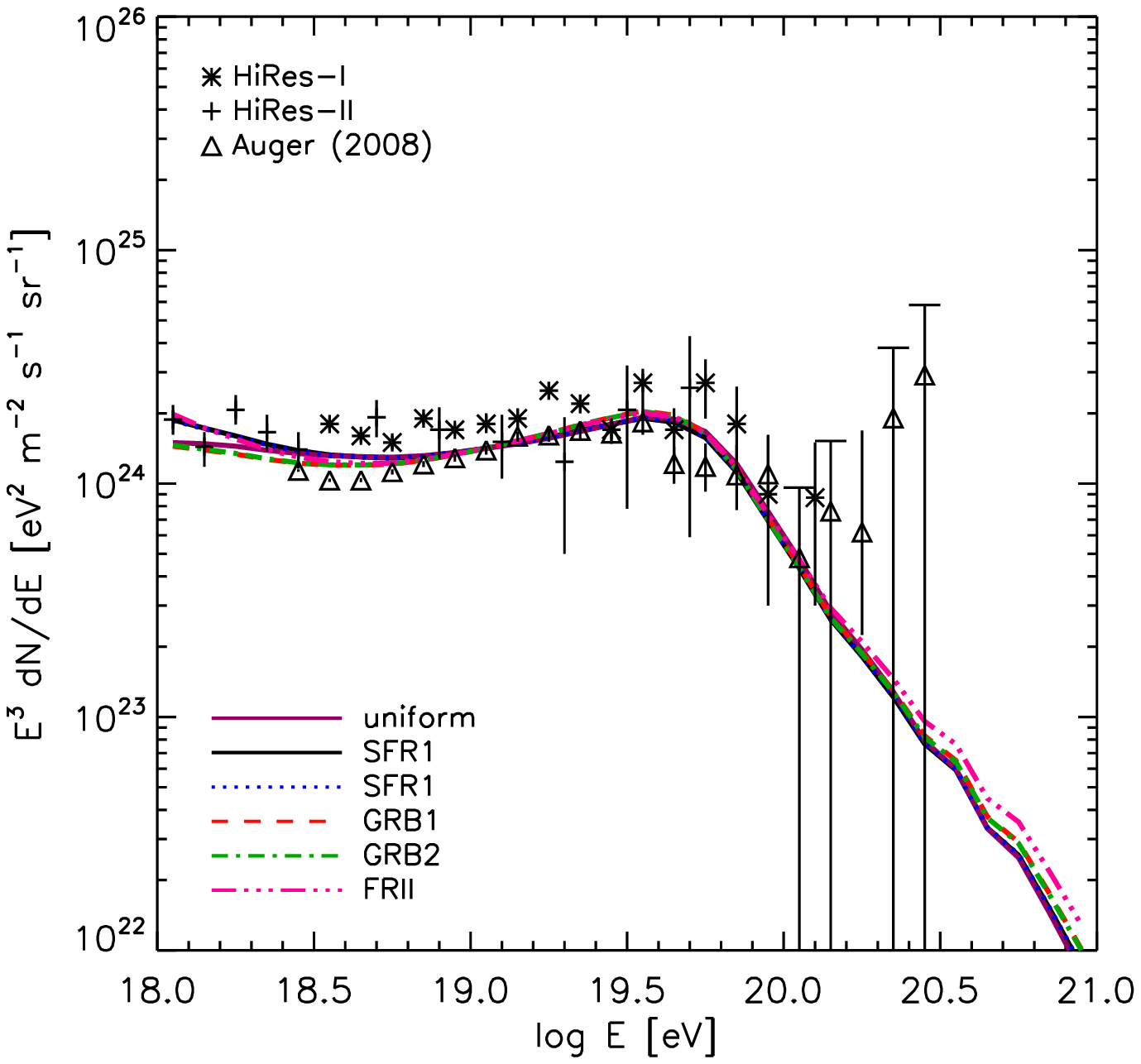} 
\includegraphics[width=0.6\textwidth]{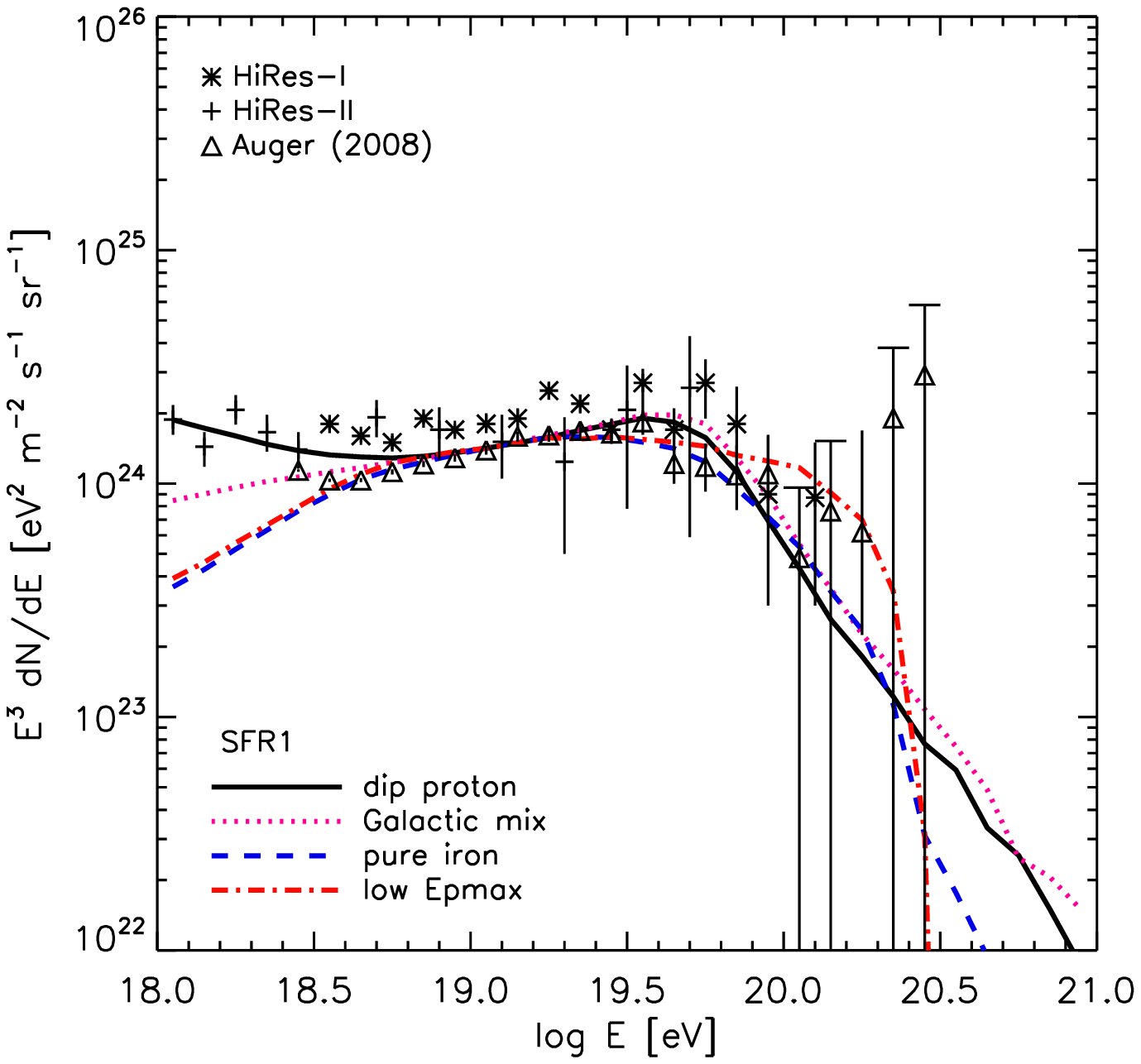} 
\caption{Propagated cosmic ray spectra adjusted to the observational data by HiRes-I and HiRes-II \citep{Abbasi04} and the Auger Observatory \citep{Auger08}. Top:  for different source evolutions, assuming a pure proton composition and a dip transition model. Bottom: for different compositions as labeled in the legend and with the corresponding spectral indices listed in table~\ref{table:indices}, assuming a SFR1 type source evolution.}  \label{fig:crspectra}
\end{center}
\end{figure}

\begin{figure*}[htb]
\begin{center}
\includegraphics[width=0.7\textwidth]{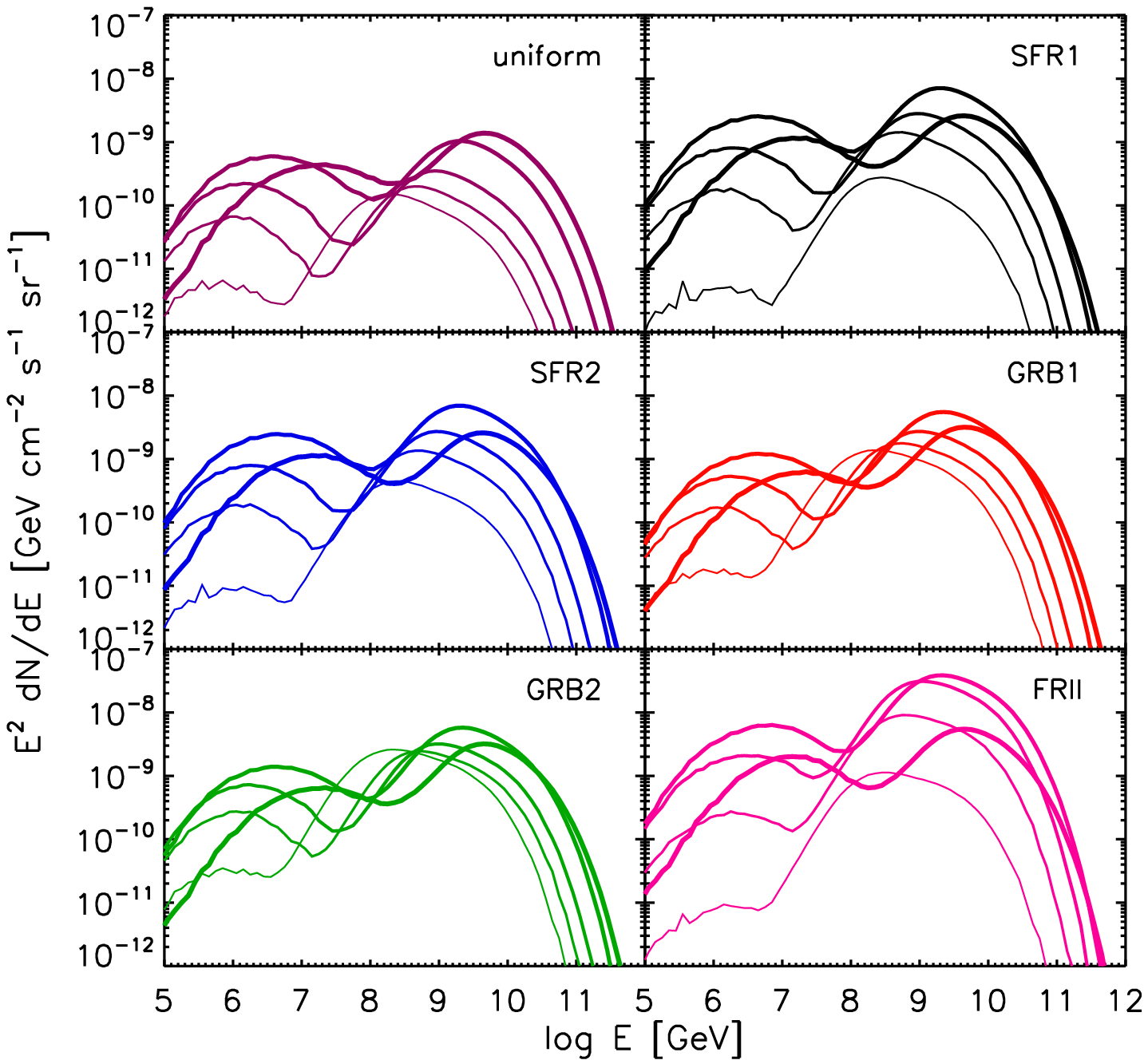} 
\caption{Contribution to the total neutrino flux of sources in different redshift bins. Each panel represents the fluxes obtained for one of the source evolution model described in the text. Each line (for increasing thickness) corresponds to the sum of the fluxes produced by sources located in the following redshift bins: $z<0.5$, $0.5\le z<1.5$, $1.5\le z<2.5$, $2.5\le z<4$ and $z>4$.}  \label{fig:redshifts}
\end{center}
\end{figure*}

Several observation-based estimates of the evolving star formation rate have been made in recent years, mostly by measuring the evolution of the galaxy luminosity functions over a broad range of wavelengths. Results from SDSS, GALEX, COMBO17, and Spitzer now allow a tighter constraint on the cosmic star formation history up to redshifts of $z\sim 1$ (see \citealp{Coward07} for a review). Above $z\sim 1$, the results of the different studies disagree on the shape of the SFR, though a tendency towards a plateau in the range of $z\sim 1-4$ and then a more or less steep decrease for $z\gtrsim 4$ seems to emerge \citep{HB06,Li08,Ota08,Yuksel08,WD09}. 

Such an evolution indicates that the cosmic photon background, especially in the UV range, is notably amplified between redshifts $z\sim 0-2$. The cosmic ray mean free path of interaction with the IR/UV background will consequently evolve with the redshift. 
The CMB photon density also increases with redshift in $(1+z)^3$, implying that the high energy bump will also be affected by the source emissivity evolution. Note that the IR/UV background evolves less than the CMB because unlike the latter it is continuously produced during the cosmic history. The decrease of this background with redshift is thus slower than the one of the CMB. The effect of the evolution is actually smaller in the IR/UV region than in the CMB region. Nevertheless, the difference in the steepness of the injected spectral indices required to adjust the propagated cosmic ray spectrum induces large variations between the fluxes at low energy.

Not many astrophysical objects fulfill the stringent energetic requirements to be potential sources of ultrahigh energy cosmic rays. 
The main candidate sources are the following: transient sources such as gamma ray bursts (GRB) or young magnetars, and continuous sources like powerful active galactic nuclei (AGN). Among AGN, Faranoff-Riley type I (FRI) and II (FRII) galaxies are more specifically discussed, though FRI galaxies are far from satisfying the energetic criteria to accelerate particles to the highest energies (see \citealp{LW09}). It might be worth mentioning as well that no outstanding correlation has been observed between catalogues of FRII galaxies and the most energetic events seen by Auger, which does not give strong credence to these types of sources, unless the particle rigidity is unexpectedly low. 

To describe the redshift evolution of the emissivity of these candidate sources, we use the latest measurements and studies made on these objects. The star formation rate (SFR) being a general tracer of matter density in the Universe, and thus a (possibly biased) tracer of ultrahigh energy cosmic ray sources, we also examine the effect of two recently derived SFR trends.  Star forming galaxies in particular might host transient objects such as GRB and young magnetars. 

In this paper, we model the source evolution using six typical trends (see Fig.~\ref{fig:source_evol}): 
\begin{itemize}
\item uniform: the source emissivity experiences no evolution. \cite{Beckmann03} for example argue that this might be the case for FRI type galaxies.
\item SFR1: the source evolution follows the star formation rate derived in \cite{HB06}. In this model, the source emissivity increases as $(1+z)^{3.4}$ for $z<1$, then $(1+z)^{-0.26}$ for $1\le z<4$ and $(1+z)^{-7.8}$ for $z\ge 4$.
\item SFR2: the source evolution follows the star formation rate from \cite{Yuksel08}. In this model, the source emissivity increases as $(1+z)^{3.4}$ for $z<1$, then $(1+z)^{-0.3}$ for $1\le z<4$ and $(1+z)^{-3.5}$ for $z\ge 4$.
\item GRB1 and GRB2:  the latest {\it Swift} GRB observations indicate that the GRB rate departs from the SFR at the highest redshifts \citep{DRM06, LD07,GP07,WD09}. We chose two models of GRB rate evolution that follow closely the SFR up to $z\sim 4$ and then continues to increase with a more or less shallow slope. The source emissivity evolves as $(1+8z)/[1+(z/3)^{1.3}]$ for GRB1 and as $(1+11z)/[1+(z/3)^{0.5}]$ for GRB2, as derived in \cite{LD07}. 
\item FRII: recent measurements indicate that FRII type galaxies follow a steep evolution for $z<4$ (\citealp{Wall05}, see also \citealp{HSM05}). We compute the emissivity evolution $\dot{\rho}$ according to \cite{Wall05} as $\log \dot{\rho} = 2.7z+1.45z^2+0.18z^3-0.01z^4$. \cite{HSM05} argue that FRI type galaxies might follow a similar, though less steep trend.
\end{itemize}

Figure~\ref{fig:source_evol} presents the cosmogenic neutrino fluxes for all flavors, obtained for these source evolutions, for a pure proton dip model case. 
We checked that the following discussion remains identical for all other composition and transition models. The normalization of the fluxes was calculated by adjusting our propagated ultrahigh energy cosmic ray spectra to the observational data, as shown in Fig.~\ref{fig:crspectra}. The cosmic ray spectra obtained for the different source evolution models match well the data for the chosen spectral indices (see Table~\ref{table:indices}) and normalizations. Figure~\ref{fig:crspectra} shows how the UHECR spectra data alone cannot specify the source model. 

Figure~\ref{fig:source_evol} shows that the source evolution introduces an overall scaling of the neutrino fluxes. The fact that the neutrino fluxes obtained for SFR1, SFR2, GRB1 and GRB2 models are very close demonstrates that the difference of evolution above $z\sim 4$ has a minor impact on the neutrino production. Indeed, the maximum differences in flux come from the combined effects of a sharp increase of the source emissivity and of the sharp decrease of the mean free path to interaction with the CMB and the IR/UV background, at redshifts $z\lesssim2$. The contribution of sources at redshifts $z\gtrsim4$ represents less than 1\% of the total flux due to redshift  dilution. These effects are illustrated in Fig.~\ref{fig:redshifts}, where the contributions of sources at various redshifts are shown. Let us note that these fluxes, unlike the case of UHE photons (see for instance the recent paper by \citealp{TA09}), would not be strongly affected by a possible overdensity or underdensity of sources in the local universe.

\subsection{Effects of various transition models}\label{subsection:transition}

\begin{figure}[thb]
\begin{center}
\includegraphics[width=0.6\textwidth]{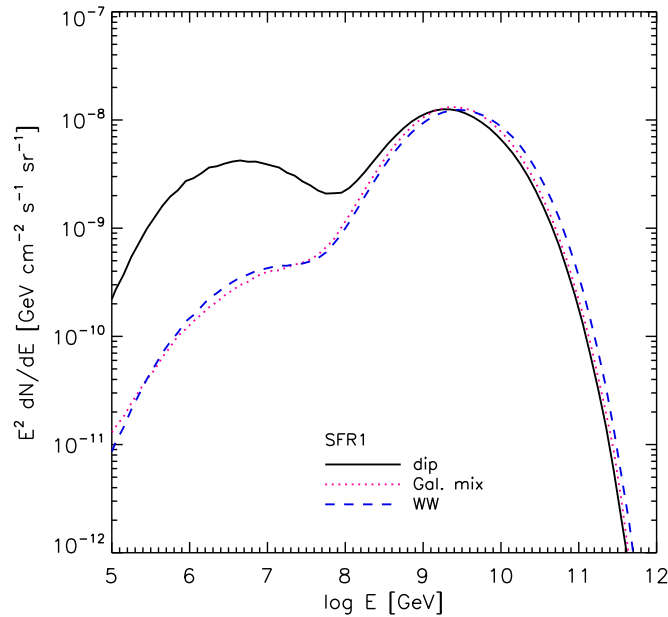} 
\caption{Effects of various transition models on neutrino fluxes for all flavors. We present the case of a source evolution following the star formation rate from \cite{HB06}. Black solid line: the pure proton `dip model' with an injection spectrum of 2.5, pink dotted: transition slightly below the ankle for a Galactic mixed composition with an injection spectrum of 2.1, blue dashed line: pure proton `WW model' with a transition at energy $> 10^{19}~$eV with a 2.1 injection spectrum (see text for description of models). }  \label{fig:transition}
\end{center}
\end{figure}

Around energies $E\sim10^{18.5-19}~$eV, the cosmic ray spectrum hardens, creating a feature commonly referred to as the `ankle'. In the standard picture, the ankle is associated with the transition between the Galactic and the extragalactic components. Early versions of transition models proposed that the extragalactic cosmic rays emerge at very high energy ($E> 10^{19}~$eV) and be composed of 100\% protons, as the measurements seemed to indicate at that time (see e.g. \citealp{W95}). \cite{WW04} follow this idea and fit the shape of the ankle by injecting particles with spectral index $2.0-2.4$ at the source. Their extragalactic component begins to predominate over the Galactic one above $E\sim 4\times10^{19}$~eV. The injected composition is highly enriched in protons.
Throughout this paper, we will refer to this ankle transition model, most recently developed by \cite{WW04} as the `WW model'.

For the mixed chemical composition model (for which the extragalactic cosmic ray composition at the source is assumed to be similar to that of low energy Galactic cosmic rays), \cite{Allard05} demonstrated that the shape of the spectrum can be well reproduced, assuming an injection spectrum $\alpha$ of order $2.2-2.3$. 
In this model, the transition between Galactic and extragalactic components happens at lower energy ($E\sim$~EeV) and ends at the ankle.

\cite{BGG06} proposed that this transition occurs at even lower energy, around $E\sim 10^{16.5-17.5}$~eV, where the cosmic ray spectrum may steepen, creating the so-called `second knee'. The combination of the second knee and the ankle is viewed in this model as a dip due to pair production energy losses during the intergalactic propagation. This scenario eases the issue of particle acceleration up to high energy inside the Galaxy, that is raised by the other models. It requires however a relatively steep injection spectrum ($2.3-2.7$ according to the assumed source evolution) that can induce an energy budget problem for extragalactic sources if the power-law remains identical down to the energy of the second knee. This problem can be bypassed by assuming a broken power-law at injection \citep{BGG06}. Again in this scenario, heavy elements cannot exceed 10-15\% of the total composition in order to fit the shape of the spectrum \citep{BGG06,Allard07}. This model will be referred to as the `dip model'.

Figure~\ref{fig:transition} presents the effects of these three transition models on the cosmogenic neutrino flux, assuming that the source evolution follows the star formation rate from \cite{HB06}. The spectral indices at injection needed to fit the observational cosmic ray data differ according to the chosen model (see Table~\ref{table:indices}). The hardening of the spectrum is responsible for the amplitude of the low energy bump in the neutrino flux. The difference of flux observed in Fig.~\ref{fig:transition} around $1-10$~PeV is directly related to the number of extragalactic cosmic rays present at low energy.
At the highest energy end of the cosmic ray spectrum, the effect of the harder spectral injection needed for the WW model is counterbalanced by the fact that the extragalactic component does not account for the whole cosmic ray flux at $10$~EeV, energy at which we normalize our calculated flux. As a result the flux is similar to what is found for the dip model. In the mixed composition model, although the cosmic ray flux is purely extragalactic above the ankle, the gain of the harder spectral index (compared to the dip model), is compensated by the presence of nuclei (that produce less neutrinos than protons for a 2.1 spectral index) in the source composition. For these reasons, the neutrino fluxes expected at the highest energies are almost identical for the three transition models although the WW and mixed composition models require harder spectral indices. 

The weak dependence of neutrino fluxes on transition models at EeV energies was highlighted by \cite{Takami09}. We see in the following that the maximum acceleration energy and the composition can affect the neutrino flux in this energy range significantly. 

\subsection{Effects of various maximum acceleration energy}\label{subsection:epmax}
\begin{figure}[tb]
\begin{center}
\includegraphics[width=0.6\textwidth]{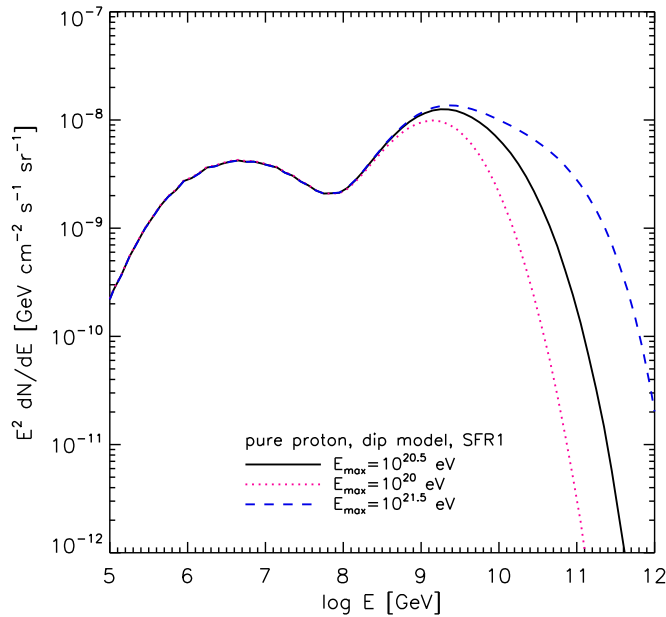} 
\caption{Effects of various maximum acceleration energy for protons $E_{p,{\rm max}}$ on neutrino fluxes for all flavors. We present here the case of a pure proton dip transition model (see section~\ref{subsection:transition} for description of the model), assuming a SFR1 type source evolution. $E_{p,{\rm max}}= 10^{20}, 10^{20.5}$ and $10^{21}$~eV for respectively: pink dotted, black solid and blue dashed lines.}  \label{fig:epmax}
\end{center}
\end{figure}

The maximal energy at which particles can be accelerated depends on the source energetics and on numerous physical parameters of the acceleration site. For a given source model, the maximum energy can in principle be estimated by comparing the cosmic ray acceleration time to the escape time, the source lifetime, and energy loss times due to various processes like interaction and dynamical expansion (see, e.g., \citealp{Norman95,AP09}). Since acceleration sites are not yet known the physical processes that may reach the highest energies are far from clear. Moreover, one expects that $E_{p,\rm max}$ varies among the sources (see, e.g.,  \citealp{KS06}).

We present in Fig.~\ref{fig:epmax} the effects of three different maximum acceleration energy on the shape of the neutrino flux. We present the case of a pure proton dip transition model, assuming a SFR1 type source evolution. This figure demonstrates the robustness of the high energy neutrino peak around $E_\nu\sim 10^{18-18.5}$~eV, with respect to the maximum acceleration energy. The main difference in flux occurs at $E_\nu \gtrsim 10^{19}$~eV, which corresponds to the energy range covered by ANITA and JEM-EUSO. In the next section, we discuss however that a too low $E_{p,\rm max}$, that would fall below the proton photo-pion production threshold, can lead to a drastic suppression of the neutrino flux, especially around $\sim$ EeV energies.

\subsection{Effects of various compositions}\label{subsection:composition}

The chemical composition of ultrahigh energy cosmic rays remains an open question. Measurements prior to the Pierre Auger Observatory indicated an increasingly lighter composition above $E\sim 10^{17}$~eV \citep{Fly, AGASA, Abu-Zayyad00,Hires,Hires10}. The latest results of the Pierre Auger Observatory suggest a mixed composition at all energies, that gets heavier at the highest end \citep{Auger_Xmax10}. Furthermore, there is no reliable theoretical prediction of the expected composition at the source, mainly because very little is known about the physical parameters that govern the acceleration and survival of nuclei in those powerful objects. 

We thus consider in this study four typical compositions that have been shown to fit the shape of the observed ultrahigh energy spectrum: (i) a pure proton composition in the dip model case, (ii) a proton dominated mixed composition based on Galactic cosmic ray abundances as in \cite{Allard06}, (iii) a pure iron composition and (iv) a mixed composition that was proposed by \cite{Allard08}, that contains 30\% of iron. For this last model, the maximum proton energy is $E_{p,\rm max}=10^{19}~$eV and a hard injection spectrum of 1.2 in the SFR evolution case is needed in order to correctly adjust the observed cosmic ray spectral shape (see Fig.~\ref{fig:crspectra}). In this scenario, the propagated cosmic ray composition would be heavy at the highest energy, as favored by \cite{Auger_Xmax10}. For the first three models, we choose the maximum proton injection energy of $E_{p,\rm max}=10^{20.5}~$eV. In all cases, we assume $E_{Z,\rm max}=Z\times E_{p,\rm max}$ for a nucleus of charge number $Z$. We take an exponential cut-off for the spectrum.

\begin{figure}[tb]
\begin{center}
\includegraphics[width=0.6\textwidth]{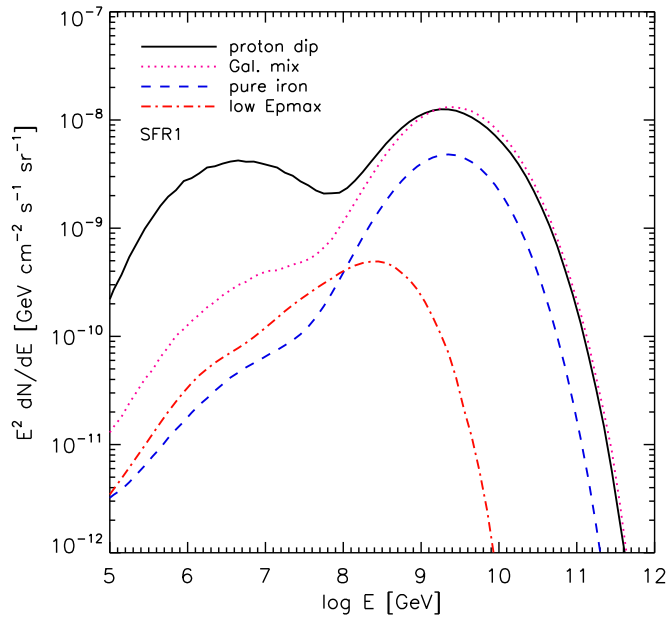} 
\caption{Effects of various compositions on neutrino fluxes for all flavors. We present the cases of (i) a pure proton injection assuming a dip transition model (black solid), (ii) a proton dominated Galactic type mixed composition (pink dotted), (iii) pure iron composition (blue dashed) and (iv) the iron rich low $E_{\rm p, max}$ model (red dash-dotted). }  \label{fig:composition}
\end{center}
\end{figure}

\begin{figure}[tb]
\begin{center}
\includegraphics[width=0.6\textwidth]{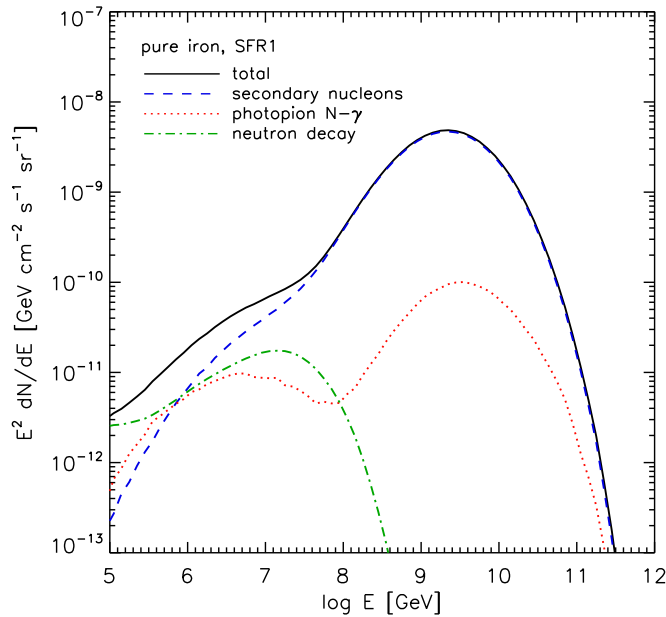} 
\caption{Contribution of different neutrino production channels to the total cosmogenic neutrino flux for all flavors (black solid line). Case of pure iron, with injection spectral index 2.0 and assuming a SFR1 type source evolution. Blue dashed line: neutrinos produced by secondary nucleons, red dotted line: neutrinos produced directly by photo-pion process of the primary nuclei, green dash-dotted line: contribution of the decay of secondary neutrons. }  \label{fig:compo}
\end{center}
\end{figure}

Figure~\ref{fig:composition} presents the neutrino fluxes obtained in these four scenarios. It appears that the high energy peak is only mildly dependent on all composition models, except for the case of the iron rich low $E_{p,\rm max}$ scenario. Indeed, as illustrated in Fig.~\ref{fig:compo}, neutrinos are mainly produced via photopion production of secondary nucleons that are themselves produced by photo-disintegration processes of the primary nuclei. For heavy nuclei, the high energy neutrino flux thus depends on the rate of particles that have an energy per nucleon higher than the pion production threshold of protons on the CMB ($E_A/A>E_{p\gamma_{\rm CMB}}$, where $A$ is the atomic number of a nucleus of energy $E_A$). When comparing with fluxes expected in the pure proton case at a similar maximum energy per charge, the difference will ultimately depend on the spectral index needed to fit experimental data (see \citealp{Allard06} for more details). In the SFR source evolution case the spectral indices for proton and iron sources are respectively 2.5 and 2.0, resulting in a factor of 3 different in the high energy neutrino peak. In the FRII case, the spectral indexes required are respectively 2.3 and 1.6, and the two composition hypotheses result in similar neutrino fluxes at the highest energies. In the iron rich case (iv), the pion production threshold is not reached for most of the particles because of the low $E_{\rm p, max}$. The direct photopion production of primary nuclei is also suppressed because of the too low $E_{Z,{\rm max}}$. Let us note here that in our calculation we assume that sources are standard candles and all have the same maximum energy. In particular in the $E_{\rm p, max}$ all the sources have an exponential cut-off above $10^{19}$ eV (note that then, $E_{Z=26,\rm max}=2.6\times 10^{20}$~eV). This assumption could be somewhat relaxed to allow some fraction of the sources to reach higher maximum energy for protons without changing the global phenomenological feature of the low $E_{\rm p, max}$ model. The expectations for neutrino fluxes at high energies would be higher in this case but would obviously remain below these obtained for standard proton dominated models, except if the maximum energy reached by the sources is strongly correlated with the redshift. Indeed, in the latter case, if the sources at large redshift had a larger proton maximum energy, one could obtain large neutrino fluxes together with a heavy composition at the highest energy, the protons accelerated up to the highest energies being produced outside the energy losses horizon.

The strong variations in amplitude at lower energy stem from the injected spectral indices that change the number of particles available for producing $\sim1-10$~PeV energy neutrinos. Thus the low energy peak is strongly affected by changes in injected composition. 

\begin{figure*}[htb]
\begin{center}
\includegraphics[width=0.8\textwidth]{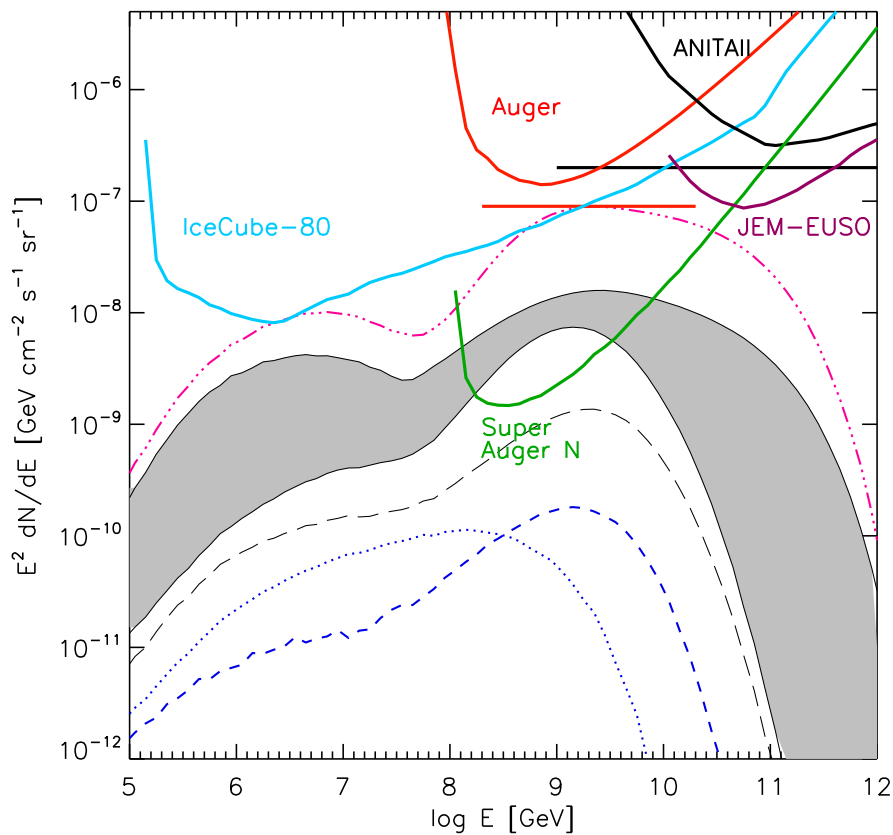} 
\caption{Cosmogenic neutrino fluxes for all flavors, for different parameters compared to various instrument sensitivities. The pink dot-dashed line corresponds to the FRII strong source evolution case with a pure proton composition, dip transition model and $E_{p,\rm max}=10^{21.5}$~eV. Blue lines are our extreme pessimistic cases: the blue dotted line represents the iron rich, low $E_{p,\rm max}$ composition, and the blue dashed line the pure iron injection case, with $E_{p,\rm max}=10^{20}$~eV;  both lines assume a uniform evolution of sources. The shaded area brackets a wide range a parameters: all transition models and all source evolutions except uniform and FRII, for pure protons and a mixed `Galactic' composition are considered. Including the uniform source evolution would broaden the shaded area down to the black long-dashed line. Instrument sensitivities: differential limits for super Auger North multiplied by 3 (green dashed, see text), IceCube 80 lines averaged over the three flavors (blue dash-dotted, acceptance from S. Yoshida, private communication, see also \citealp{Karle10} and \citealp{Abbasi10}), and JEM-EUSO multiplied by 3 (purple solid, acceptance from \citealp{JEM-EUSO}, see text). In red solid line: differential limit and integral flux limit on a pure $E^{-2}$ spectrum (straight line), both multiplied by 3 (see text) for Auger South, using the optimistic acceptance from \cite{Auger_nu09}. In black solid line: ANITA-II differential limit at 90\% CL, for 27.1 day livetime, for all flavors, the straight line indicates the integral flux limit on a pure $E^{-2}$ spectrum \citep{ANITA10}.}   \label{fig:global}
\end{center}
\end{figure*}

\begin{figure}[htb]
\begin{center}
\includegraphics[width=0.6\textwidth]{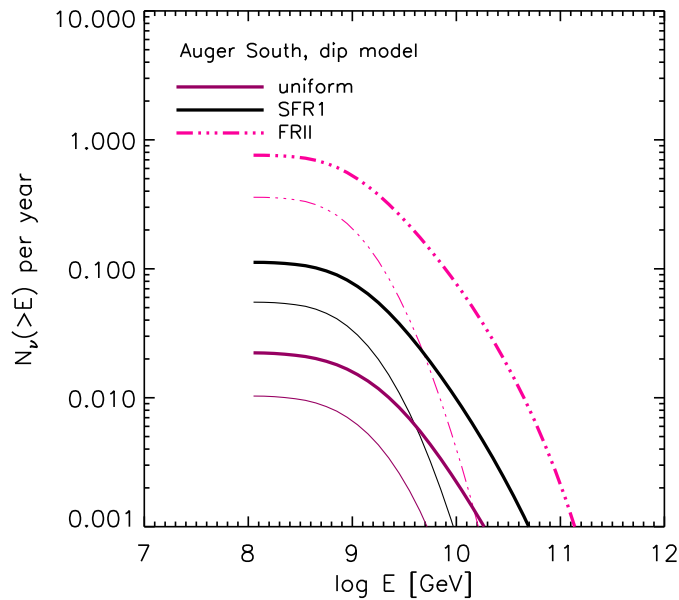} 
\caption{Cumulative number of neutrinos per year above a given energy expected for Auger South, in the `reasonable' parameter range represented by the grey shaded area in Fig.~\ref{fig:global}: thick lines for the upper bound and thin lines for the lower bound.
The numbers are calculated using the optimistic exposures given in \cite{Auger_nu09}.  }  \label{fig:numAuger}
\end{center}
\end{figure}

\begin{figure}[htb]
\begin{center}
\includegraphics[width=0.6\textwidth]{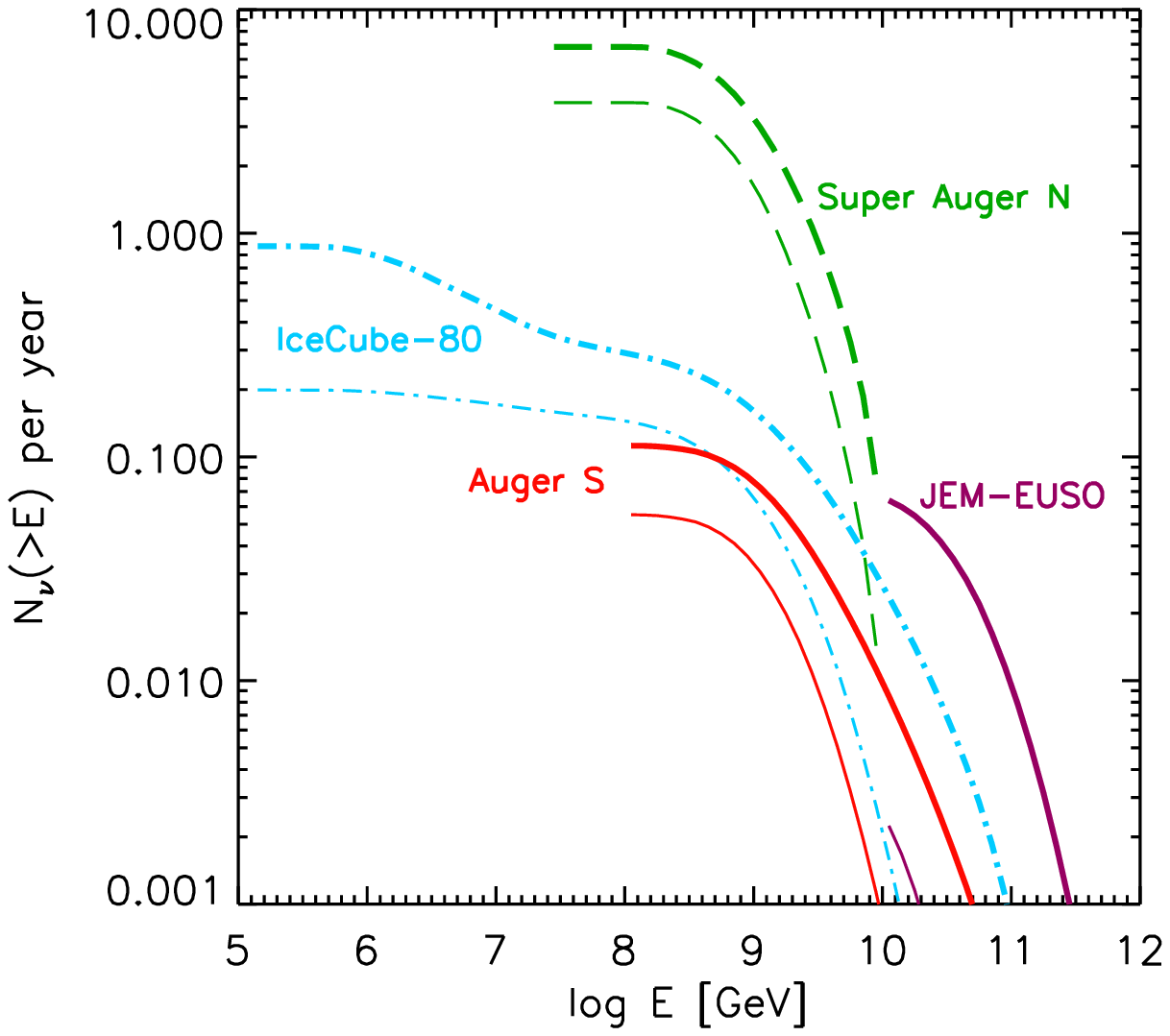} 
\caption{Cumulative number of neutrinos (all flavors) per year above a given energy expected for Auger South (red solid, optimistic acceptance from \citealp{Auger_nu09}), super Auger North (green dashed, see text), IceCube 80 lines for three flavors (blue dash-dotted, acceptance from S. Yoshida, private communication, see also \citealp{Karle10}), and JEM-EUSO (purple solid, acceptance from \citealp{JEM-EUSO}).  The numbers are calculated for the `reasonable' parameter range represented by the grey shaded area in Fig.~\ref{fig:global}: thick lines for the upper bound and thin lines for the lower bound.}  \label{fig:numAll}
\end{center}
\end{figure}

\section{Implications for the existing and upcoming detectors}\label{section:detection}

Figure~\ref{fig:global} summarizes our results and compares our fluxes to the existing, upcoming, and possible future neutrino detector sensitivities. Our estimates for neutrino fluxes are divided into three possible regions: an optimistic scenario (pink dot-dashed line), a plausible range of models in which we base many of our rate estimates (grey shaded area), and a more pessimistic scenario (blue lines). The optimistic scenario corresponds to the FRII strong source evolution case with a pure proton composition, dip transition model and $E_{p,\rm max}=10^{21.5}$~eV. The most pessimistic scenario is given by a pure iron injection and the iron rich composition with low $E_{p,\rm max}$, assuming in both cases a uniform evolution of sources. The shaded area brackets a wide range of parameters: all discussed transition models,  all source evolutions except for uniform and FRII, and varying cosmic ray injection composition from pure protons to a mixed Galactic type model, with $E_{p,\rm max}\ge 10^{20}$~eV. The black long-dashed line indicates the minimum neutrino flux one could obtain in the case of a uniform source evolution, when the composition and the maximum acceleration energy are chosen among reasonable values. Namely, this line represents the case of a Galactic mixed composition with $E_{p, \rm max}=10^{20}~$eV for a uniform source evolution. 

From the discussion elaborated at the beginning of section~\ref{subsection:source_evol}, it stands out that a uniform UHECR source evolution should be deemed rather extreme. Indeed, under the assumption that UHECRs are produced in astrophysical sources, the majority of their plausible progenitors should follow -- with a possible bias -- the star formation history. Though \cite{Beckmann03} suggest that FRI-type galaxies might have experienced a quasi-uniform emissivity evolution throughout time, one should be aware that these radio-galaxies are, as already discussed, very bad candidates for UHECR production due to their poor energetics. As for the FRII source evolution, we chose to consider it as an extreme scenario as well, and not to include it in our `reasonable' grey shaded area. This optimistic scenario from the neutrino flux point of view is not favored by UHECR observations due to the lack of correlations between the FRII galaxies and the highest energy events seen by Auger.

Figure~\ref{fig:global} reveals that the cosmogenic neutrino flux can vary of many orders of magnitude throughout the whole energy range. In the PeV energy region, the full sets of models imply a three order of magnitude variation while one has an uncertainty of four orders of magnitude in the EeV region. The ZeV region is unbound from below, making a detection in this range the most speculative. If one focuses on the `reasonable' domain (the grey shaded area in Figure~\ref{fig:global}), it appears that the spread around EeV energies becomes fairly limited in regard to the various parameters. The one exception to this robust behavior is a very low $E_{p,\rm max}$. In the PeV region, our `reasonable' models introduce a wider span in flux, which can be helpful to distinguish among the various models, if neutrinos are found in both energy ranges.

In terms of detectability, one may first note that the optimistic scenario is currently being constrained by observations. 
For more plausible sets of parameters (grey region), the EeV energy range is close to the sensitivity of the Auger Observatory and of IceCube-80 (see also the numbers in Figs.~\ref{fig:numAuger} and~\ref{fig:numAll}). At PeV energies, the situation is more uncertain due to the high variation of the flux according to the parameters. It should be noted however that IceCube-80 will start to detect neutrinos or at least give interesting constraints in this region very soon. ZeV neutrino observatories, such as ANITA and JEM-EUSO will constrain the maximum acceleration energy in the most optimistic case. The sensitivity in the ZeV energy range will however have to be greatly improved in order to explore this parameter space.

The green line in Fig.~\ref{fig:global} labeled `Super Auger North' represents the sensitivity of an EeV neutrino detector covering the area proposed for Auger North (20,000~km$^2$, \citealp{whitepaper,Bluemer10}). If Auger North had 100\% detection efficiency for neutrino showers around an EeV (as assumed in this plot), for example, through a denser array than currently proposed, then its sensitivity would be vastly superior to currently planned observatories in this energy range. An alternative to a denser array of Auger surface detectors is the possibility of new atmospheric air-shower detection techniques such as radio or microwave.

Figures~\ref{fig:numAuger} and \ref{fig:numAll} give the cumulative numbers of neutrinos that are expected for the different instruments, in the case of a pure proton composition, dip transition model, with SFR1 type source evolution, which roughly corresponds to the upper limit of the shaded area in Fig.~\ref{fig:global}. The numbers are only represented in the instrument sensitivity range. From these figures, one can infer that at EeV energies, one should detect $0.06-0.2$ neutrino per year with IceCube-80 and $0.03-0.06$ with Auger South. Let us note however that current and planned experiments should be unable to detect cosmogenic neutrino fluxes predicted for the dip model if there is no evolution of the cosmic ray luminosity with redshift. 
In point of fact, the {\it non} observation of cosmogenic neutrinos in the next few years would certainly help us constrain UHECR source evolution models (see e.g. \citealp{Berezinsky09}). On the other hand, it would not be possible to constrain the source composition or the Galactic to extragalactic transition models without positive detection over the next decades.

Interestingly, it was pointed out by recent studies \citep{Ahlers10,Berezinsky10} that the strongest source evolution models can be constrained also by the cascading of cosmogenic photons down to GeV energies. These authors have shown that (at least for proton dominated compositions up to the highest energies) the Fermi-LAT diffuse gamma-ray flux would be overshot by the products of cascading UHE photons, for scenarios where the comoving source luminosity is very strong. Although the two studies cited above do not agree on the implications of this constraint on the expectations for cosmogenic neutrino fluxes, one can say that our FRII source evolution model with a pure proton composition certainly fits in this particular case.\\

Throughout this paper, we have always plotted the neutrino fluxes for all flavors. The comparison of these total fluxes to instrument sensitivities requires an assumption on the proportions of flavors. If cosmogenic neutrinos are produced with a ratio $\nu_e:\nu_\mu:\nu_\tau=1:2:0$, subsequent neutrino oscillation during the propagation should lead to a proportion on Earth of $1:1:1$ \citep{Learned95}. We will not consider here other neutrino mixing behaviors than the canonical mixing described for example in \cite{Pakvasa08}.

\begin{figure}[htb]
\begin{center}
\includegraphics[width=0.6\textwidth]{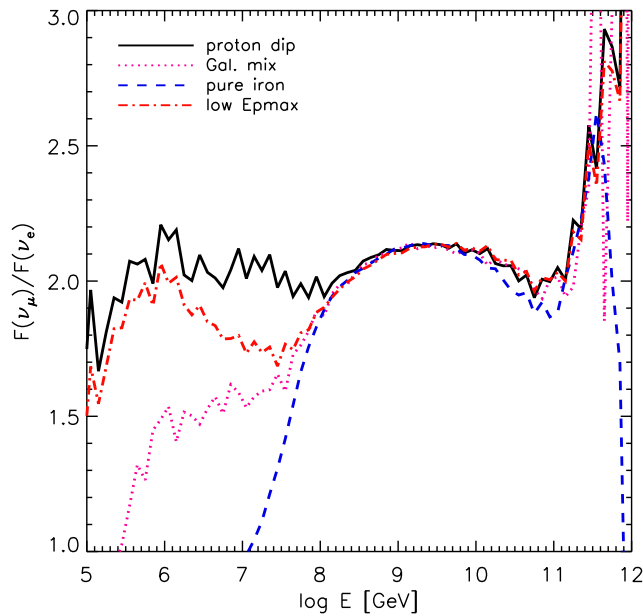} 
\caption{Ratio of fluxes of $\nu_\mu$ to $\nu_e$ produced during the propagation of UHECRs for the various injected chemical compositions described in section~\ref{subsection:composition}. We present the cases of (i) a pure proton injection assuming a dip transition model (black solid), (ii) a proton dominated Galactic type mixed composition (pink dotted), (iii) pure iron composition (blue dashed) and (iv) the iron rich low $E_{\rm p, max}$ model (red dash-dotted). }  
\label{fig:flavors}
\end{center}
\end{figure}

Figure~\ref{fig:flavors} shows that these proportions are indeed reached in the case of a pure proton composition, in the range of energy that is of interest for our study. For the other chemical compositions, the contribution of the neutron decay at $E_\nu<10^8$~GeV (see Fig.~\ref{fig:compo} for the iron case) enhances strongly the proportion of electronic neutrinos. For the injection with a low $E_{p,\rm max}$, the production of secondary neutrons is lower due to the low overall energy of nuclei and this effect is attenuated. In Fig.~\ref{fig:global}, we represent the sensitivity of IceCube averaged over the three neutrino flavors. The fact that the ratio between $\nu_\mu$ and $\nu_e$ departs from 2 for non pure proton compositions indicates that at PeV energies, the detection of $\nu_e$ can be favored against other flavors. The ratios are of $\sim1:3/2:0$ for a Galactic composition, which should lead after oscillation to only a slight excess for $\nu_e$: $\sim 1.28:1:1$, and the detection for a pure iron composition is highly compromised in any case. Due to the slight difference between the sensitivities of each flavor (within a factor $2-3$) for IceCube, we took into account the individual flavor fluxes and acceptances when calculating the number of expected neutrinos in Fig.~\ref{fig:numAll}.

At EeV energies on the other hand, the ratio remains stable around $1:2:0$. For this reason, we multiplied in Fig.~\ref{fig:global} the sensitivities for $\tau$ neutrinos of Auger South and Super Auger North by 3, assuming a final proportion at the Earth of $1:1:1$. We applied the same multiplication to the JEM-EUSO sensitivity, though the assumption of equal ratios may seem debatable. Indeed, as can be seen in Fig.~\ref{fig:flavors}, at the highest energies ($E_\nu>10^{11}$~GeV), the production of muonic neutrinos is enhanced due to the decay of Kaons and other mesons that lead to different neutrino spectra than for muon decay. One can calculate however that the neutrino oscillation should lead in principle to flavor proportions of order $0.88:1:1$ for flux ratios $F(\nu_\mu)/F(\nu_e)\sim 3$, which remains close to the standard $1:1:1$ ratio. Our calculations do not consider the muon polarization effects that should modify the $\nu_\mu$ to $\nu_e$ flavor ratio at the highest energies as was demonstrated by \cite{Lipari07}. This issue should again affect only slightly the overall expected $\nu_\tau$ flux for JEM-EUSO.
We checked that the ratios presented in Fig.~\ref{fig:flavors} are similar for our various source evolution models.

Let us further note that in this study, we only take into account the cosmogenic neutrinos, i.e. the neutrinos produced by interactions with the radiative backgrounds during the propagation of the UHECRs in the intergalactic medium. Therefore, the issues of changes in neutrino flavor proportions due to the source environment, through damping of muon decay for instance (see e.g. \citealp{Kashti05}) do not apply here.\\

In our calculations, we neglected the effect of the extragalactic magnetic fields on the neutrino fluxes. At the highest energies ($E_{\rm cr}>10^{19}~$eV), their influence is indeed negligible unless their intensity becomes high enough to produce a sizable magnetic horizon effect at the energies where we normalize our neutrino flux to the cosmic ray data. This could happen if the mean magnetic field is $\langle B\rangle \gtrsim 10$~nG, which is not favored by current observations (see e.g. \citealp{KL08} and \citealp{Globus08}). 

It should be also remarked that the flux around PeV energies could be further amplified by more than one order of magnitude provided that most sources are located in strongly magnetized environments such as clusters of galaxies, if one stipulates that the radiation density in these environments is stronger than in the extragalactic medium. The amplitude of this effect depends on various assumptions on the source environment as is discussed in detail in \cite{KAM09}. 

Finally, because we normalized our neutrino fluxes to the observed ultrahigh energy cosmic ray spectrum \cite{Auger08}, a possible shift in energy of the cosmic ray flux would consequently change the neutrino flux. The current systematic uncertainty on the Auger spectrum energy scale being of order 30\%, one should consider that the fluxes calculated in this study bear an uncertainty of order $\sim50$\%. Note that normalizing to the AGASA data \citep{Hayashida97} would increase furthermore the neutrino fluxes of a factor 2.

Generally our cosmogenic neutrino fluxes are consistent with those found by previous authors \citep[for a review]{ESS01,Ave05,Allard06,Takami09,AM09}, provided that the spectral indices and composition are adjusted, and taking into account the differences in normalizations according to various considered cosmic ray detectors. 

It is important to underline that many figures in the literature which summarize the range of cosmogenic neutrino fluxes often also include the production of neutrinos at the acceleration site {\it inside} the source (e.g. \citealp{Chen09,Allison09,ANITA10}). Such a production is independent of the flux of cosmogenic neutrinos that are produced outside the source during the propagation in the extragalactic medium. Its detailed study requires assumptions on the opacity of the acceleration region and strongly depends on the shape of the injection spectrum as well as on the phenomenological modelling of acceleration. In the same token, recently, \cite{AAS09} claimed to obtain a `guaranteed' neutrino flux (assumed to be produced by each cosmic-ray proton escaping from the sources)  as well as stringent constraints on the proton fraction in the cosmic ray composition using AMANDA-II and IceCube limits. One should be aware however that such estimates strongly depend on the physical modeling of particle acceleration and escape from the source. Especially if the source is optically thick, cosmic rays should not be accelerated up to the highest energies as demonstrated by \cite{AP09}, and EeV energy neutrinos would be sharply suppressed. 

In optimistic source scenarios, it was calculated that the cosmogenic neutrino fluxes would be overwhelmed by the neutrinos from the acceleration site (see for example \citealp{MPR01} for an AGN scenario and \citealp{MN06} for a GRB scenario). It is interesting to note, as discussed by \cite{Takami09} that neutrinos from GRB would only overwhelm the PeV peak of the cosmogenic neutrinos and furthermore should be distinguishable, as they should be temporally and spatially correlated with with the prompt GRB emission. Again, these estimates are optimistic cases and subject to high variability according to the parameters assumed for the source. 

\section{Conclusion}
\label{section:conclusion}

We calculated the flux of neutrinos generated by the propagation of ultrahigh energy cosmic rays over cosmological distances,  using a complete numerical Monte Carlo method that takes into account the interactions of nuclei with cosmic background radiations. 

We explored the influence of different cosmic ray source evolutions, transition models, chemical compositions, and maximum acceleration energies on the cosmogenic neutrino flux -- our results are summarized in Fig.~\ref{fig:global}. We showed that the parameter space is currently poorly constrained with uncertainties of several orders of magnitude in the predicted flux. We defined three main regions in this wide parameter space: an optimistic scenario that is currently being constrained, a `reasonable' range of cosmic ray parameters, and a pessimistic low maximum proton acceleration energy where fluxes are too low to be detected in the foreseeable future.

In the `reasonable' window of the parameter space, the neutrino flux in the EeV energy range ($10^{18-19}$~eV) is quite robust in regard to these various cosmic ray parameters. We find that the normalization of the flux in this region ultimately depends on the source evolution up to redshift $z\sim 4$ and on the rate of cosmic rays with energy per nucleon higher than the pion production threshold for protons on the cosmic microwave background ($E_A/A>E_{p\gamma_{\rm CMB}}$, where $A$ is the atomic number of a nucleus of energy $E_A$). Composition models and Galactic to extragalactic transition scenarios are not uniquely determined by measurements above $10^{17.5}$~eV. These unknowns can only be constrained by complementary measurements of cosmogenic neutrinos in different energy ranges.

In our `reasonable' neutrino flux region, IceCube should observe $0.06-0.2$ neutrino per year and the Pierre Auger Observatory $0.03-0.06$ neutrino per year in the EeV range, and detection should happen in the next decade unless $E_{p, \rm max} < 6\times 10^{19}~$eV. If EeV neutrinos are detected, PeV information can help select between competing models of cosmic ray composition at the highest energy and the Galactic to extragalactic transition at ankle energies.  With improved sensitivity, ZeV neutrino observatories, such as ANITA and JEM-EUSO could explore and place limits to the maximum proton acceleration energy. 

\section*{Acknowledgments}
We thank Bruny Baret, Dariusz Gora, Aya Ishihara, Keiichi Mase, Teresa Montaruli, Kohta Murase, Markus Roth, V\'eronique Van Elewyck, and Shigeru Yoshida for providing us with the acceptances of IceCube and Auger and for very helpul discussions. KK and AVO are supported by the NSF grant PHY-0758017 and by Kavli Institute for Cosmological Physics at the University of Chicago through grant NSF PHY-0551142 and an endowment from the Kavli Foundation. AVO acknowledges the support from Agence Nationale de Recherche in France and DA was supported by C.N.R.S. in France and by NSF grant PHY-0758017 and by Kavli Institute for Cosmological Physics at the University of Chicago.

\bibliography{neutrinos}

\end{document}